\documentclass[11pt]{article}
\usepackage{biometrics_woojung}
\usepackage{enumitem}
\usepackage{natbib}
\usepackage{bm}
\usepackage{tikz}
\usepackage{soul}
\usepackage{booktabs}
\usetikzlibrary{arrows.meta, positioning, shapes.geometric}
\usepackage[ruled,vlined]{algorithm2e}
\DeclareMathOperator{\logit}{logit}
\DeclareMathOperator{\expit}{expit}
\usepackage{changes}

\title{Estimator-Aligned Prospective Sample Size Determination for Designs Using Inverse Probability of Treatment Weighting}

\author[1]{Taekwon Hong}
\author[2]{Daeyoung Lim}
\author[3]{Woojung Bae}
\author[4]{Yong Ma}

\affil[1,4]{Division of Biometrics VII, Office of Biostatistics, Office of Translational Sciences, Center for Drug Evaluation and Research, U.S. Food and Drug Administration, Silver Spring, Maryland 20993}
\affil[2]{Division of Biometrics III, Office of Biostatistics, Office of Translational Sciences, Center for Drug Evaluation and Research, U.S. Food and Drug Administration, Silver Spring, Maryland 20993}
\affil[3]{Division of Biostatistics, Office of Biostatistics and Pharmacovigilance, Center for Biologics Evaluation and Research, U.S. Food and Drug Administration, Silver Spring, Maryland 20993}

\makeatletter

\begin{document}

\maketitle

\setlength{\parskip}{2pt}
\setlength{\abovedisplayskip}{2pt}
\setlength{\belowdisplayskip}{2pt}
\setlength{\abovedisplayshortskip}{2pt}
\setlength{\belowdisplayshortskip}{2pt}


\begin{abstract}
{    
In observational studies, accurately characterizing variance is critical for sample size determination, yet unaccounted-for variability from propensity score estimation and the resulting weights limit the accuracy of standard variance approximations for design. Existing approaches often rely on heuristics or randomized controlled trial (RCT) formulas that treat weights as fixed, potentially misaligning prospective design with the causal estimator used at analysis. We propose an estimator-aligned framework for prospective sample size determination based on generalized estimating equations (GEE) and stacked M-estimation. By merging the propensity score model and marginal structural model (MSM) into a single system of estimating equations, the method propagates nuisance-model uncertainty and directly targets the large-sample variance of the IPTW estimator. For study planning, we estimate a pilot-based large-sample variance factor and introduce a bootstrap stabilization procedure that accounts for both within- and between-pilot variability. The framework applies uniformly across binary, count, and continuous outcomes through link-specific GEE representations under a common design principle. Simulation studies motivated by post-marketing safety and healthcare cost applications demonstrate that anchoring design to this variance improves power calibration relative to conventional RCT-style formulas, particularly in settings with weight instability, outcome sparsity, or heavy-tailed variability.
}
\end{abstract}


\newpage
\section{Introduction}
In randomized controlled trials (RCTs), prospective study design is guided by well-established principles \citep{ichE9,ichE9R1} that align the estimand, primary analysis, and sample size calculation based on study objectives and prior information. In observational studies that rely on inverse probability of treatment weighting (IPTW) to balance baseline covariates across treatment groups \citep[e.g.,][]{rosenbaum1983central,austin2015moving}, prospective planning often defaults to conventional RCT sample size formulas \citep[e.g.,][]{chow2017sample}. It ignores that IPTW induces data-dependent weights and weight-driven heteroskedasticity complicates variance estimation.

A growing methodological literature has therefore sought to address this challenge \citep[e.g.,][]{jung2007note, austin2021informing, shook2022power, liu2025sample}. \citet{austin2021informing} proposes diagnostics based on effective sample size and the empirical distribution of inverse probability weights to quantify precision loss induced by IPTW. Building on this idea, \citet{shook2022power} develop a design-effect framework using Kish-type approximations to inflate correct variances. \citet{liu2025sample} derive analytical expressions by decomposing the variance into components capturing covariate overlap, potential outcome variability, and their correlation.

While these approaches provide valuable insight into variability under IPTW, a few design-level limitations remain. First, power and sample size are typically formulated using variance decompositions or surrogate quantities rather than the asymptotic variance of the estimator defined by the planned analysis (e.g., \cite{shook2022power,liu2025sample}). Second, the study design is not fully aligned with the estimand-defined analysis method based on estimating equations, so key dependencies between propensity score estimation, weight instability, and marginal structural model (MSM) inference are indirectly captured, typically through summary statistics rather than direct modeling (e.g., \cite{jung2007note,shook2022power}). Specifying the propensity score model in isolation from study planning treats causal adjustment as a downstream analytic concern rather than a design-stage decision \citep{austin2021informing}. In practice, the choice of a propensity score model---including covariate selection and the functional form---directly impacts the variability of the causal estimator and hence the required sample size. 

In this paper, we propose a framework for determining sample size in IPTW analyses that aligns with the estimator used in the final analysis. The approach is based on generalized estimating equations (GEE) and stacked M-estimation \citep{stefanski2002calculus}, and it applies to binary, count, and continuous outcomes. By treating the propensity score model and the MSM as a single system of estimating equations, the stacked sandwich variance directly targets the asymptotic distribution of the treatment effect estimate under the final analysis \citep{lunceford2004stratification}. Designing studies based on this variance yields a method that is both principled and implementable. 

The proposed framework handles outcome types through appropriate link functions and working variances within the GEE formulation. For study planning, we use pilot data to estimate the variance of the causal estimator under the prespecified analysis and bootstrapping to propagate pilot-sample uncertainty into sample size calculation. The specified propensity score model at the design stage determines the estimator's variance and, consequently, the required sample size. The rest of the paper is organized as follows: Section~\ref{sec:methods} presents the methods; Sections~\ref{sec:cs_binary}--\ref{sec:cs_continuous} provide simulated case studies for binary, count, and continuous outcomes; and Section~\ref{sec:discussion} concludes.

\section{Proposed Methods}
\label{sec:methods}

\subsection{Causal Framework, Estimand, and Identification}
\label{sec:methods:causal}

Let $O_i=(X_i,T_i,Y_i)$ for $i=1,\ldots,n$ denote independent and identically distributed (i.i.d.) realized observations, where $X_i$ is a $p$-dimensional vector of baseline covariates, $T_i\in\{0,1\}$ is the treatment indicator, and $Y_i$ is the observed outcome. Under the potential outcomes framework \citep{rubin1974estimating}, let $Y_i(1)$ and $Y_i(0)$ denote the potential outcomes under treatment and control, respectively.

We target the marginal (population-average) treatment effect, i.e., the average treatment effect (ATE), modeled via the MSM with a link function $g(\cdot)$ appropriate for the outcome type:
\begin{equation}
g\!\left\{ \mathbb{E}\!\left[Y_i(t)\right] \right\}
=
\beta_0+\beta_1 t,
\qquad t\in\{0,1\},
\label{eq:msm}
\end{equation}
where $\boldsymbol{\beta}=(\beta_0,\beta_1)^\top$ is the MSM parameter vector. Under~\eqref{eq:msm}, $\beta_1$ represents the marginal treatment effect on the link-function scale. For example,
\begin{itemize}[nosep]
\item \textbf{Binary outcome (logit link):} $\beta_1=\logit\mathbb{E}[Y(1)]-\logit\mathbb{E}[Y(0)]$, the logarithm of the marginal odds ratio (OR).
\item \textbf{Count outcome (log link):} $\beta_1=\log\mathbb{E}[Y(1)]-\log\mathbb{E}[Y(0)]$, the logarithm of the incidence-rate ratio (IRR).
\item \textbf{Continuous outcome (identity link):} $\beta_1=\mathbb{E}[Y(1)]-\mathbb{E}[Y(0)]$, the mean difference.
\end{itemize}

Let $e(x):=\Pr(T=1\mid X=x)$ denote the propensity score. The following standard assumptions are sufficient for the identification of $\boldsymbol{\beta}$ from observed data---that is, $\boldsymbol{\beta}$ can be expressed as a function of observable quantities.

\begin{assumption}[Stable Unit Treatment Value Assumption (SUTVA)]
\label{ass:sutva_m}
There is no interference between study units and no hidden versions of treatment. In particular, $Y_i(t)$ depends only on $t$ and not on other subjects' treatments.
\end{assumption}

\begin{assumption}[Consistency]
\label{ass:consistency_m}
For each $i$, $Y_i=Y_i(1)$ if $T_i=1$ and $Y_i=Y_i(0)$ if $T_i=0$.
\end{assumption}

\begin{assumption}[Positivity]
\label{ass:positivity_m}
For almost every $x$ in the support of $X$, the propensity score $e(x)$ satisfies $0<e(x)<1$.
\end{assumption}

\begin{assumption}[Conditional Exchangeability]
\label{ass:exchangeability_m}
$\{Y(1),Y(0)\}\perp T \mid X$.
\end{assumption}
Assumptions~\ref{ass:sutva_m}--\ref{ass:exchangeability_m} jointly ensure identification of the MSM parameter $\boldsymbol{\beta}$ from observed data. Together, Assumptions \ref{ass:sutva_m} and \ref{ass:consistency_m} guarantee that potential outcomes are well-defined and the observed outcome corresponds to the potential outcome under the received treatment. Assumption \ref{ass:positivity_m} requires that both treatment options remain possible across the range of covariates. Assumption \ref{ass:exchangeability_m} allows the treatment effect to be identified through covariate adjustment when all relevant covariates are available \citep{rosenbaum1983central,cole2009consistency,westreich2010positivity}.

Under these assumptions, $\boldsymbol{\beta}$ is identified from observed data and can be consistently estimated via IPTW
\begin{equation}
w_i(\eta)
=
\frac{T_i}{e(X_i;\eta)}+\frac{1-T_i}{1-e(X_i;\eta)},
\label{eq:iptw_weight}
\end{equation}
where $e(X_i;\eta)=\Pr(T_i=1\mid X_i;\eta)$ is a parametric propensity score model indexed by nuisance parameter $\eta\in\mathbb{R}^{p_\eta}$. Alternative causal estimands, such as the average treatment effect among the treated (ATT), can be accommodated by modifying the weight function accordingly---e.g., $w_i^{'}(\eta)=T_i+(1-T_i)\,e(X_i;\eta)\{1-e(X_i;\eta)\}^{-1}$ for the ATT. For simplicity, we focus on the ATE formulation in this paper without loss of generality.

\subsection{Stacked Estimating Equations for Joint Nuisance-Structural Inference}
\label{sec:methods:stacked}

We specify a parametric propensity score model $e(X;\eta)=\Pr(T=1\mid X;\eta)$. Let $U_{i,\eta}(\eta)\in\mathbb{R}^{p_\eta}$ denote the score vector for $\eta$ corresponding to the $i$-th patient:
\begin{equation*}
\sum_{i=1}^n U_{i,\eta}(\widehat\eta)=0.
\end{equation*}
For example, under logistic regression, $e_i(\eta)=\mathrm{logit}^{-1}(X_i^\top\eta)$ and
$U_{i,\eta}(\eta)=X_i\{T_i-e_i(\eta)\}$.

Let $\mu(t;\boldsymbol{\beta})=g^{-1}(\beta_0+\beta_1 t)$ denote the marginal mean under the working MSM. The corresponding weighted estimating equation is
\begin{equation}
U_{i,\boldsymbol{\beta}}(\boldsymbol{\beta},\eta)
=
D_{i}\, w_i(\eta)\,\{Y_i-\mu_i(\boldsymbol{\beta})\},
\qquad
\sum_{i=1}^n U_{i,\boldsymbol{\beta}}(\widehat{\boldsymbol{\beta}},\widehat\eta)=0,
\label{eq:msm_score_glmform}
\end{equation}
where $D_{i}$ denotes the design matrix defining the MSM. The choice of $D_{i}$ may incorporate effect-modification terms or alternative link-mean structures. In what follows, we assume $D_{i}=(1,T_i)^\top$. Define the stacked estimating
function
\[
U_i(\theta)=
\begin{pmatrix}
U_{i,\eta}(\eta)\\[2pt]
U_{i,\boldsymbol{\beta}}(\boldsymbol{\beta},\eta)
\end{pmatrix},
\qquad
\theta=(\eta^\top,\boldsymbol{\beta}^\top)^\top\in\mathbb{R}^{p_\eta+2}.
\]
The estimator $\widehat\theta$ is $\theta$ such that
$\sum_{i=1}^n U_i(\widehat\theta)=0$.

\subsection{Large-Sample Variance of the Stacked Estimator}
\label{sec:methods:variance}

Under standard regularity conditions for M-estimation \citep{stefanski2002calculus, lunceford2004stratification}, $\widehat\theta$ is asymptotically normal. That is,
\begin{equation*}
\sqrt{n}\,(\widehat\theta-\theta_0)
\ \xrightarrow{d}\
N\!\left(0,\ \Sigma\right),
\qquad
\Sigma = A^{-1} B (A^{-1})^\top,
\end{equation*}
where $\theta_0$ denotes the true parameter value, $A = \mathbb{E}\!\left[\nabla_{\theta} U_i(\theta_0)\right]$ for which $\nabla_\theta = \partial/\partial \theta^\top$, and $B = \mathbb{E}\!\left[U_i(\theta_0)U_i(\theta_0)^\top\right]$. 
The corresponding empirical sandwich estimator is $\widehat{\Sigma}
=
\widehat{A}^{-1}\widehat{B}\left(\widehat{A}^{-1}\right)^\top$, where
\begin{equation}
\widehat{A}
=
\frac{1}{n}\sum_{i=1}^n
\nabla_\theta U_i(\widehat\theta),
\qquad
\widehat{B}
=
\frac{1}{n}\sum_{i=1}^n
U_i(\widehat\theta)U_i(\widehat\theta)^\top.
\label{eq:empirical_sandwich}
\end{equation}
For IPTW estimators, these conditions additionally require smoothness of the propensity score model and sufficient overlap to ensure finite weight moments (Assumption~\ref{ass:positivity_m}). In Equation~\eqref{eq:msm_score_glmform}, we adopt a working-independence specification and use the standard GEE score representation
$D_{i}\,w_i(\eta)\{Y_i-\mu(T_i;\boldsymbol{\beta})\}$.
Inference is based on the empirical sandwich estimator, which remains statistically consistent even if the working variance is misspecified, and ensures alignment between design and analysis.
Alternative working variances can be incorporated by redefining $U_{i,\boldsymbol{\beta}}$, provided that pilot and final analyses use identical estimating equations. 

Let $\Sigma_{\beta}$ denote the $2\times 2$ submatrix of $\Sigma$ corresponding to $\beta$, and similarly let $\widehat{\Sigma}_{\beta}$ be the same submatrix of $\widehat{\Sigma}$. Then, $\mathrm{Var}(\widehat{\beta}_1)\approx \Sigma_{\beta, 22}/n$ and $\widehat{\mathrm{Var}}(\widehat{\beta}_1)\approx \widehat{\Sigma}_{\beta,22}/n$. Because the propensity score estimating equations do not depend on $\boldsymbol{\beta}$, the Jacobian matrix $A$ has a block lower-triangular structure:
\begin{equation}
A=
\begin{pmatrix}
A_{\eta\eta} & 0 \\
A_{\beta\eta} & A_{\beta\beta}
\end{pmatrix},
\label{eq:block_A}
\end{equation}
where $A_{\beta\eta}=\mathbb{E}\!\left[\partial U_{i,\beta}(\beta_0,\eta_0)/\partial\eta^\top\right]$ captures the sensitivity of the MSM estimating equation to perturbations in the propensity score parameters and characterizes how uncertainty in nuisance-parameter estimation propagates to the variance of $\widehat{\boldsymbol{\beta}}$.

\subsection{Prospective sample size from the LSVF target}
\label{sec:methods:samplesize}

We consider a two-sided Wald test for $H_0:\beta_1=0$ versus $H_1:\beta_1\neq 0$,
where $\beta_1$ is the MSM treatment parameter on the link scale and $\Delta\neq 0$ is a clinically meaningful alternative on that same scale.
The test rejects when
\[
\left|\frac{\widehat\beta_1}{\widehat{\mathrm{SE}}(\widehat\beta_1)}\right|>z_{1-\alpha/2},
\]
where $\alpha$ is the significance level and $1-\gamma$ is power. Under $H_1:\beta_1=\Delta$ and a normal approximation for $\widehat{\beta}_1$, targeting power $1-\gamma$ approximately yields the standard requirement
\begin{equation}
\mathrm{SE}(\widehat\beta_1)=\frac{|\Delta|}{z_{1-\alpha/2}+z_{1-\gamma}}.
\label{eq:se_target}
\end{equation}
Let $V$ denote the large-sample variance factor (LSVF) such that $\mathrm{Var}(\widehat\beta_1)\approx V/n$. Conceptually, $V$ represents the contribution of a single observation to the large-sample variance.
Then,
\begin{equation}
n=\frac{(z_{1-\alpha/2}+z_{1-\gamma})^2}{\Delta^2}\,V.
\label{eq:n_from_V}
\end{equation}

\subsection{Pilot-based estimation and bootstrap stabilization of $V$}
\label{sec:methods:pilot_boot}

Let $\mathcal{D}_{\mathrm{pilot}}$ denote pilot data of size $n_{\mathrm{pilot}}$ drawn from the target population under the intended covariates and treatment definition.
Applying the stacked estimating procedure to $\mathcal{D}_{\mathrm{pilot}}$ yields
$\widehat{\mathrm{Var}}_{\mathrm{pilot}}(\widehat\beta_1)$ via the $\beta$-block of~\eqref{eq:empirical_sandwich}, and we form
\[
\widehat{V}_{\mathrm{pilot}}=n_{\mathrm{pilot}}\widehat{\mathrm{Var}}_{\mathrm{pilot}}(\widehat\beta_1).
\]

The design-relevant quantity, however, is not a single realization of $V$, but rather its distribution across repeated samples from the target population, reflecting variability in covariate distributions, treatment prevalence, and weight stability under estimated propensity scores. A single pilot dataset provides only one draw from this distribution and therefore does not capture between-trial variability.

To approximate this distribution, we generate $B$ bootstrap resamples
$\{\mathcal{D}_{\mathrm{pilot}}^{*(b)}\}_{b=1}^B$ and compute
\[
\widehat{V}^{*(b)}=
n_{\mathrm{pilot}}\widehat{\mathrm{Var}}^{*(b)}(\widehat\beta_1),
\qquad b=1,\ldots,B.
\]
The empirical distribution $\{\widehat V^{*(b)}\}_{b=1}^B$ serves as a proxy for the sampling distribution of $V$ under the design, capturing variability induced by finite-sample estimation.

We then select a design value via a stability functional $\mathcal{F}$,
\[
\widehat{V}_{\mathrm{stable}}=\mathcal{F}\!\left(\widehat{V}^{*(1)},\ldots,\widehat{V}^{*(B)}\right),
\]
interpreted as a risk-sensitive decision rule. In particular, taking $\mathcal{F}$ as an upper empirical quantile $Q_q(\cdot)$ (e.g., $q \in \{0.5, 0.7, 0.9 \})$ yields a data-adaptive conservative estimate that robustifies variance estimation, whereas the mean corresponds to a risk-neutral choice.

To further account for the sampling variability of the functional $\mathcal{F}$ itself, we construct a bootstrap-based upper confidence bound (UCB) for a generic functional $\phi$ of the bootstrap distribution (not necessarily identical to $\mathcal{F}$) \citep[e.g.,][]{peter2003ucb,hao2019bootucb}. Specifically, we resample $B_\text{ucb}$ times from $\{\widehat V^{*(b)}\}_{b=1}^B$; for each resample $k$, let $\phi^{*(k)}=\phi(\widehat{V}^{*\pi(1)}, \ldots, \widehat{V}^{*\pi(B)})$, $k=1,\ldots,B_{\mathrm{ucb}}$, where $\pi_k(\cdot)$ denotes sampling with replacement from $\{1, \ldots, B\}$. We then define
\[
\widehat{V}_{\mathrm{stable}}^{\mathrm{UCB}}
=
Q_{1-\gamma_{\mathrm{ucb}}}\!\left(\phi^{*(1)},\ldots,\phi^{*(B_{\mathrm{ucb}})}\right).
\]
This construction yields an UCB for the chosen functional $\phi$, inflating the design variance in proportion to the instability of the pilot-induced bootstrap distribution. Consequently, a greater dispersion in $\{\widehat{V}^{*(b)} \}$ leads to more conservative design values, providing a data-adaptive mechanism for variance inflation and power assurance. The second bootstrap is computationally negligible---and optional---as it operates only on the scalar quantities $\{\widehat{V}^{*(b)}\}_{b=1}^B$. In general, these stability functionals can be viewed as operational design rules that explicitly trade efficiency against robustness to pilot uncertainty---
pseudocode and conceptual illustrations are provided in Supplementary Appendices~\ref{app:S1}~and~\ref{app:S2}.

\subsection{Stabilized design and empirical power calibration}
\label{sec:methods:design_validation}
\begin{figure}[t]
\centering
\begin{tikzpicture}[
  node distance=9mm and 14mm,
  box/.style={draw, rounded corners, align=center, inner sep=6pt, font=\small},
  bigbox/.style={box, minimum width=0.9\linewidth},
  arrow/.style={-{Latex[length=2.2mm]}, thick}
]

\node[bigbox] (pilot) {%
\textbf{Pilot generation}\\
Observed pilot data $\mathcal{D}_{\mathrm{pilot}}$ or synthetic pilot sampled under the target design\\
(Size $n_{\mathrm{pilot}}$; covariates $X$, treatment $T$, outcome $Y$)
};

\node[bigbox, below=of pilot] (fit) {%
\textbf{Pilot analysis via stacked estimating equations}\\
Estimate propensity score $e(X;\widehat\eta)$ and fit MSM for $\beta$ using IPTW\\
Obtain $\widehat{\mathrm{Var}}_{\mathrm{pilot}}(\widehat\beta_1)$ from stacked sandwich
};

\node[bigbox, below=of fit] (unit) {%
\textbf{Pilot LSVF}\\
$\widehat{V}_{\mathrm{pilot}} = n_{\mathrm{pilot}}\cdot
\widehat{\mathrm{Var}}_{\mathrm{pilot}}(\widehat\beta_1)$
};

\node[bigbox, below=of unit] (boot) {%
\textbf{Bootstrap stabilization}\\
Bootstrap $\mathcal{D}_{\mathrm{pilot}}^{*(b)}$, $b=1,\ldots,B$\\
Compute $\widehat{V}^{*(b)}$ and summarize via $\mathcal{F}$
(quantile, mean, or UCB)
};

\node[bigbox, below=of boot] (design) {%
\textbf{Prospective sample size}\\
$n_{\mathrm{prop}}=
\dfrac{(z_{1-\alpha/2}+z_{1-\gamma})^2}{\Delta^2}\,
\widehat{V}_{\mathrm{stable}}$
};

\node[bigbox, below=of design] (validate) {%
\textbf{Empirical power validation}\\
Repeat pilot-based design across $R$ pilots\\
Estimate interpolated power via Monte Carlo simulation over a sparse grid of $n$
};

\draw[arrow] (pilot) -- (fit);
\draw[arrow] (fit) -- (unit);
\draw[arrow] (unit) -- (boot);
\draw[arrow] (boot) -- (design);
\draw[arrow] (design) -- (validate);

\end{tikzpicture}
\caption{Pilot-based prospective sample size determination for IPTW marginal structural models.
The workflow combines stacked M-estimation, bootstrap-stabilized LSVF,
and empirical power validation under the target alternative.}
\label{fig:workflow}
\end{figure}
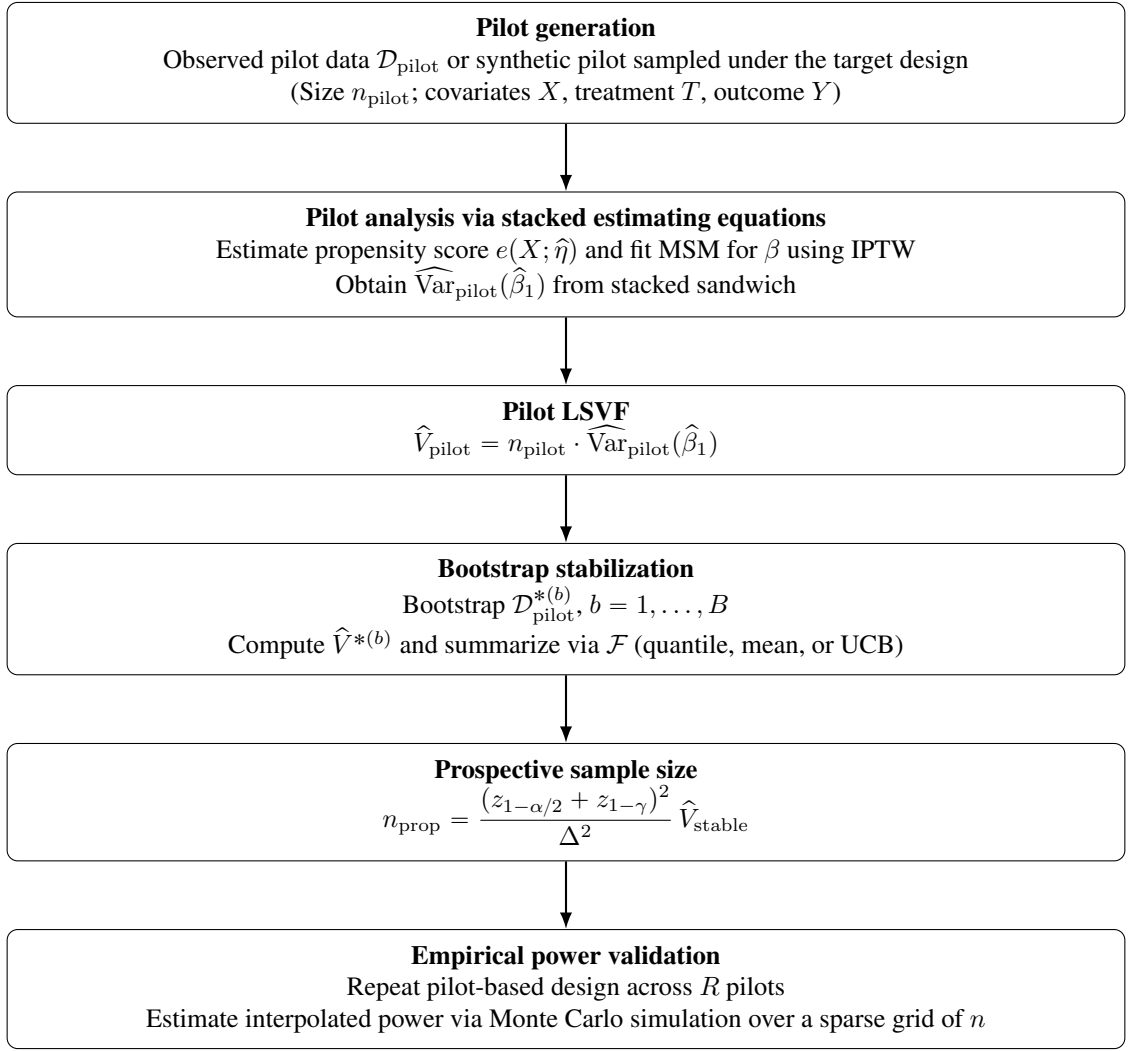

To investigate the performance of the proposed method, we repeatedly simulate independent pilot datasets under controlled settings and apply the full design procedure to each dataset, yielding a collection of proposed stabilized sample sizes $\{n_{\mathrm{prop}}^{(r)}\}_{r=1}^R$, where 
$n_{\mathrm{prop}} = \widehat{V}_{\mathrm{stable}}(z_{1-\alpha/2}+z_{1-\gamma})^2/\Delta^2.$
Although the data-generating mechanism is known, closed-form analytic power for a given proposed sample size $n_{\mathrm{prop}}$ is generally unavailable because the asymptotic variance depends on the sandwich covariance $\Sigma = A^{-1} B (A^{-1})^\top$, whose components involve population expectations over $(Y,T,X)$ and nonlinear functions of the estimated propensity score and induced weights. We therefore evaluate empirical power via additional Monte Carlo simulation step under the intended stacked-sandwich Wald test.

Because power evaluation relies on simulation, assessing every integer value of $n$ is computationally infeasible. Instead, we pool the $\{n_{\mathrm{prop}}^{(r)}\}_{r=1}^R$ values across stability functionals to construct a sparse grid of candidate sample sizes, including selected quantiles, the mean, the minimum and maximum, and the RCT-style benchmark given a data generation procedure. This grid spans the practically relevant design range while maintaining computational tractability and enabling interpolation between grid points. Figure~\ref{fig:workflow} summarizes the overall workflow.

\section{Case Study I: Post-Marketing Pregnancy Safety (Binary)}
\label{sec:cs_binary}

To motivate our simulation setting, we consider a class of postmarketing safety studies with common design challenges. These evaluations often involve binary clinical endpoints, rely on real-world data from electronic health records, and feature imbalanced exposure groups relative to comparators. Such characteristics complicate prospective design especially when confounding adjustment is required. A typical example arises in studies assessing the safety of prescription opioids during early pregnancy, where first-trimester exposure is examined for potential associations with major congenital malformations (MCM) or other pregnancy outcomes \citep[e.g.,][]{broussard2011maternal,lind2017maternal}. In these observational settings, confounding adjustment is essential because opioid exposure is correlated with maternal comorbidities, concomitant medications, and healthcare utilization \citep{bateman2021association}. Our simulation case study is designed to reflect these key statistical features.

\subsection{Simulation settings}\label{subsec:simulation-settings}

For data generation, we specify two models: one for treatment assignment and one for potential outcomes. We generate i.i.d.\ observations with a baseline confounder $X\sim N(0,1)$, a binary treatment $T\in\{0,1\}$, and a binary outcome $Y\in\{0,1\}$. 
\begin{itemize}[nosep]
    \item Treatment assignment follows a logistic regression model:
\[
\Pr(T=1\mid X=x)=\expit(\eta_0+\eta_1 x), \qquad \eta_1=0.8,
\]
where $\eta_0$ is calibrated within each dataset to achieve $\Pr(T=1)\approx 0.25$.
\item The potential outcomes follow the logistic model for $T=t$:
\[
\Pr\{Y(t)=1\mid X=x\}=\expit(\gamma_0+\beta_x x+\psi t), \qquad \beta_x=0.5,
\]
with $\gamma_0$ calibrated so that $\mathbb{E}\{Y(0)\}=p_0$. Given the noncollapsibility of the logistic link, $\psi$ does not equal the marginal log-odds ratio. We therefore numerically calibrate $\psi$ to satisfy the MSM contrast:
\[
\logit\{\mathbb{E}(Y(1))\}-\logit\{\mathbb{E}(Y(0))\}=\Delta, \qquad \Delta=\log(2).
\]
\end{itemize}

We then consider two scenarios with binary outcomes: (i) a low event-rate safety endpoint (event rate $p_0=0.03$; e.g., MCM \citep{centers2008update}) and (ii) a secondary endpoint with a higher event rate ($p_0=0.10$; e.g., small for gestational age).

For each scenario, we repeat the pilot-based design procedure across $R=1000$ independently generated pilot datasets of size $n_{\mathrm{pilot}}=600$. Within each pilot, we compute the bootstrap distribution of LSVF estimates using $B=1000$ resamples, $\{\widehat V^{*(b)}\}_{b=1}^B$, and summarize it by stability functionals
$\mathcal{F}\in\left\{Q_{0.5},\,Q_{0.7},\,Q_{0.9},\,\mathrm{mean}\right\}$,
together with UCB variants obtained from a second-level bootstrap of size $B_{\mathrm{ucb}}=1000$ and tail probability $\gamma_{\mathrm{ucb}}=0.05$ (with $p_{\mathrm{ucb}}=0.5$ for the UCB-quantile functional).

We estimate empirical power under the target alternative using Monte Carlo simulation (2000 replicates per grid point). For each scenario, power is evaluated on a sparse grid of $n$ values derived from the pooled $\{n_{\mathrm{prop}}^{(r)}\}_{r=1}^R$ distribution and summarized by the hit rate $\Pr\{\widehat{\pi}(n_{\mathrm{prop}}^{(r)}) \ge 0.80\}$, where $\widehat{\pi}(\cdot)$ denotes the interpolated empirical power. We report the hit rate as a measure of design reliability across heterogeneous pilot realizations. Unlike mean power, which summarizes average performance at a fixed $n$, the hit rate characterizes the frequency with which the induced design attains the nominal power target under the intended analysis. No universal benchmark for acceptable hit rates exists; their interpretation is inherently context-dependent, reflecting study objectives, tolerance for underpowering risk, and efficiency considerations. In settings where weight instability varies substantially across samples, mean power alone can obscure a nontrivial risk of underpowered designs for a subset of plausible pilot draws. Inference for $\beta_1$ uses the stacked-sandwich variance from the joint propensity-score and MSM estimating equations, and power is evaluated using the stacked-sandwich Wald test. 

For comparison, we consider an RCT-style benchmark based on a delta-method approximation under independent sampling. The required sample size is determined by an outcome-specific variance $V_\text{RCT}$, as summarized in Table~\ref{tab:rct_variance}. The alternative parameter $\Delta$ is defined on the corresponding link scale. In the binary outcome case study, the treatment proportion $\rho$ is set to $0.25$.

\begin{table}[htbp]
\centering
\caption{RCT-style variance formulas for common outcome types. The required sample size is 
$n_{\text{RCT}} = V_{\text{RCT}}(z_{1-\alpha/2}+z_{1-\gamma})^2/\Delta^2$. We set $\alpha=0.05$ and $1-\gamma=0.8$ through the case studies.}
\label{tab:rct_variance}
\begin{tabular}{lll}
\toprule
Outcome type & Estimand (link scale) & $V_{\text{RCT}}$ \\
\midrule
Binary (log-odds ratio) 
& $\Delta = \logit(p_1) - \logit(p_0)$ 
& $\displaystyle \frac{1}{\rho\,p_1(1-p_1)} + \frac{1}{(1-\rho)\,p_0(1-p_0)}$ \\[10pt]

Count (log-incidence rate ratio) 
& $\Delta = \log(\lambda_1) - \log(\lambda_0)$ 
& $\displaystyle \frac{1}{\rho\,\lambda_1} + \frac{1}{(1-\rho)\,\lambda_0}$ \\[10pt]

Continuous (mean difference) 
& $\Delta = \mu_1 - \mu_0$ 
& $\displaystyle \frac{\sigma^2}{\rho} + \frac{\sigma^2}{1-\rho}$ \\
\bottomrule
\end{tabular}
\end{table}

\subsection{Results}

\begin{table}[t]
\centering
\caption{Binary case study results across $R=1000$ independently generated pilots ($n_{\mathrm{pilot}}=600$ per pilot; $B=1000$ pilot-level bootstraps). For each stability choice, we report the induced prospective mean and median sample size across pilots. The hit rate is defined as the proportion of pilots for which the interpolated empirical power at the induced $n$ meets or exceeds $0.80$. Empirical power curves were estimated using 2000 Monte Carlo replicates per grid point. Reported power intervals correspond to empirical 95\% confidence intervals, defined as the 2.5th and 97.5th percentiles of the empirical power distribution across pilots. Proposed sample sizes are rounded up to the nearest integer.}
\label{tab:binary_results}
\begin{tabular}{llccc}
\toprule 
Scenario & Stability choice & \shortstack{Prospective Sample Size\\ Mean (Median)} & \shortstack{Mean Power (\%) \\ (95\% CI)} & \shortstack{Hit rate (\%) \\(Power$\geq 80\%$)} \\
\midrule
\multicolumn{5}{l}{\textbf{(A) Low event rate: $p_0=0.03$}}\\
& RCT benchmark     & 1940 (1940) & 77.2 (77.2, 77.2) & 0.0\\
& Median ($Q_{0.5}$)   & 2300 (2204) & 79.1 (64.9, 95.7) & 43.8 \\
& UCB--Median       & 2337 (2240) & 79.4 (65.5, 95.9) & 46.3 \\
& Mean              & 2469 (2343) & 81.4 (66.9, 96.9) & 54.1 \\
& UCB--Mean         & 2511 (2378) & 82.3 (67.5, 97.0) & 56.6 \\
& Quantile ($Q_{0.7}$) & 2643 (2509) & 85.5 (70.5, 97.1) & 65.6 \\
& Quantile ($Q_{0.9}$) & 3371 (3129) & 90.8 (77.2, 97.9) & 89.5 \\
\midrule
\multicolumn{5}{l}{\textbf{(B) Higher event rate: $p_0=0.10$}}\\
& RCT benchmark     & 682 (682) & 75.4 (75.4, 75.4) & 0.0 \\
& Median ($Q_{0.5}$)   & 771 (761) & 79.0 (70.5, 88.8) & 38.8 \\
& UCB--Median       & 778 (768) & 79.6 (70.7, 89.1) & 42.1 \\
& Mean              & 786 (776) & 79.6 (71.0, 89.6) & 45.3 \\
& UCB--Mean         & 792 (781) & 79.6 (71.3, 90.0) & 48.5 \\
& Quantile ($Q_{0.7}$) & 830 (816) & 82.7 (72.9, 90.8) & 61.5 \\
& Quantile ($Q_{0.9}$) & 932 (904) & 86.5 (76.8, 92.1) & 84.9 \\
\bottomrule
\end{tabular}
\end{table}

Table~\ref{tab:binary_results} compares the operating characteristics of the proposed pilot-based designs with the conventional RCT-style benchmark across stability choices. The RCT benchmark, which assumes independent sampling without estimated weights, yields smaller sample sizes ($n=1940$ for $p_0 = 0.03$ and $n=683$ for $p_0 = 0.10$) reflecting its more efficient modeling assumptions. However, when evaluated under IPTW-MSM analysis, these designs are systematically underpowered, achieving only 77.2\% and 75.4\% mean power, respectively, and failing to attain the nominal 80\% target across pilot replicates, with a hit rate of 0\%. In contrast, the proposed estimator-aligned approach incorporates weighting-induced variability, leading to larger sample sizes and substantially improved power calibration.

Among the proposed designs, central stability summaries provide moderate variance adjustment but exhibit nontrivial variability in achieved power across pilot realizations. For example, median-based stabilization increases the mean sample size to 2300 in the low event-rate setting and 771 in the higher event-rate setting, yielding mean power near the target (79.1\% and 79.0\%) but relatively low hit rates (43.8\% and 38.8\%). Mean-based stabilization is more conservative, further increasing sample sizes (2469 and 786) and improving mean power (81.4\% and 79.6\%) and hit rates (54.1\% and 45.3\%), reflecting sensitivity to larger variance realizations. UCB-adjusted summaries introduce a modest additional upward shift (e.g., 2511 and 792), with corresponding gains in reliability (hit rates 56.6\% and 48.5\%) without substantial inflation of sample size.

More pronounced gains in reliability are obtained through upper-quantile stabilization. Moving to $Q_{0.7}$ increases sample sizes to 2643 and 830 and raises hit rates to 65.6\% and 61.5\%, while maintaining mean power above the nominal target (85.5\% and 82.7\%). Using the more conservative $Q_{0.9}$ further increases sample sizes (3371 and 932) and achieves high reliability (hit rates 89.5\% and 84.9\%), at the cost of elevated mean power (90.8\% and 86.5\%). All things considered, these results illustrate a continuum of design choices, where the stability functional governs the tradeoff between efficiency and robustness to variance uncertainty under the intended IPTW analysis.

\section{Case Study II: Rare Neuropsychiatric Safety Signals (Count)}
\label{sec:cs_count}

Observational studies in pharmacoepidemiology frequently encounter settings characterized by rare outcomes, count endpoints over person-time, and reliance on large healthcare databases requiring confounding adjustment. A representative example is the evaluation of neuropsychiatric events (NPEs) following influenza treatment. Oseltamivir, the predominant antiviral therapy for influenza \citep{dobson2015oseltamivir,butler2020oseltamivir,antoon2023trends}, has been the subject of longstanding safety discussions, particularly regarding serious NPEs in pediatric patients \citep{CIDRAP2005}. To examine associations between oseltamivir exposure and serious NPEs among pediatric influenza cases, \citet{antoon2025oseltamivir} analyzed 692,295 pediatric influenza episodes in Tennessee Medicaid. We conduct a simulation study designed to reflect the key statistical features of large-scale observational studies of neuropsychiatric events following influenza treatment, including rare count outcomes, person-time incidence, and confounding adjustment.

We adopt the simulation framework described in Section~\ref{subsec:simulation-settings}. The propensity score follows the same logistic regression model, with $\eta_1 = 0.5$ and treatment prevalence $P(T=1)$ calibrated to approximately $\rho=0.67$.

Following \cite{antoon2025oseltamivir}, potential outcomes are generated from Poisson model to reflect rare count outcomes:
$$
\log \lambda(t,x) = \log(\lambda_0) + \beta_x x + \Delta t,
$$
with $\beta_x = 0.3$, $\lambda_0 = 0.008$, and $\Delta = \log(0.5)$. The analysis model is a Poisson log-link MSM, so that $\beta_1=\log\{\mathbb{E}(Y\mid T=1)\}-\log\{\mathbb{E}(Y\mid T=0)\}$ represents the marginal log-incidence rate ratio. See Table~\ref{tab:rct_variance} for the RCT benchmark. We use the same simulation parameters ($R$, $B$, and Monte Carlo replicates). However, to reduce computational burden, we set the incidence level $\lambda_0$ to $0.008$ and use a relatively large pilot sample size of $n_\text{pilot}=5000$.

\subsection{Results}

\begin{table}[ht]
\centering
\caption{Count case study results across $R=1000$ independently generated pilots ($n_{\mathrm{pilot}}=5000$ per pilot; $B=1000$ pilot-level bootstraps). For each stability choice, we report the induced prospective sample size summarized as mean (median) across pilots. The hit rate is defined as the proportion of pilots for which the interpolated empirical power at the induced $n$ meets or exceeds $0.80$. Empirical power curves were estimated using 2000 Monte Carlo replicates per grid point. Reported power intervals correspond to empirical 95\% confidence intervals (2.5th and 97.5th percentiles across pilots). Proposed sample sizes are rounded up to the nearest integer.}
\label{tab:count_results}
\begin{tabular}{llccc}
\toprule 
Scenario & Stability choice & \shortstack{Prospective Sample Size\\ Mean (Median)} & \shortstack{Mean Power (\%) \\ (95\% CI)} & \shortstack{Hit Rate (\%) \\ (Power $\geq 80\%$)} \\
\midrule
\multicolumn{5}{l}{\textbf{Count outcome: $\lambda_0=0.008$}}\\
& RCT benchmark & 12284 (12284) & 76.3 (76.3, 76.3) & 0.0 \\
& Median ($Q_{0.5}$) & 14577 (14004) & 80.8 (66.8, 95.5) & 51.2 \\
& UCB--Median & 14803 (14198) & 81.4 (67.3, 96.0) & 53.4 \\
& Mean & 15564 (14763) & 82.7 (68.3, 97.1) & 58.8 \\
& UCB--Mean & 15813 (14965) & 83.2 (68.8, 97.1) & 61.2 \\
& Quantile ($Q_{0.7}$) & 16672 (15772) & 84.9 (71.4, 97.1) & 71.6 \\
& Quantile ($Q_{0.9}$) & 20821 (19356) & 90.3 (76.8, 97.6) & 92.5 \\
\bottomrule
\end{tabular}
\end{table}

Table~\ref{tab:count_results} summarizes the prospective sample size distributions induced by the pilot and the empirical power calibration for the very rare outcome setting. Consistent with Case Study~I of binary outcome, the RCT-style benchmark is misaligned with the IPTW-MSM estimator and remains overoptimistic.
Even with the RCT-based sample size of ($n_{\mathrm{RCT,count}}=12284$) attaining empirical power of $76.3\%$ on the validation grid, it fails to achieve the nominal $80\%$ target in any pilot replicate, yielding a hit rate of $0.0\%$.

Our proposed pilot-based procedure yields appropriately calibrated designs. Median-based stabilization provides moderate correction, resulting in mean power of $80.8\%$ (95\% CI: 66.8\%--95.5\%) and a hit rate of $51.2\%$. Mean-based stabilization is more conservative, increasing mean power to $82.7\%$ (95\% CI: 68.3\%--97.1\%) and improving the hit rate to $58.8\%$, consistent with greater sensitivity to occasionally large LSVF realizations. UCB-adjusted summaries introduce a modest additional upward shift (e.g., UCB--Mean: mean power $83.2\%$, 95\% CI: 68.8\%--97.1\%; hit rate $61.2\%$), reflecting pilot uncertainty without excessive sample size inflation. Upper-quantile stabilization yields the strongest calibration. Using $Q_{0.7}$ increases mean power to $84.9\%$ and the hit rate to $71.6\%$, while $Q_{0.9}$ produces substantially more conservative designs (mean power $90.3\%$) and achieves the target power in $92.5\%$ of pilot replicates.

\section{Case Study III: Post-Progression Healthcare Costs (Continuous)}
\label{sec:cs_continuous}

Projected U.S. health spending growth (5.4\%) is expected to outpace GDP growth (4.6\%) from 2022--2031, increasing the share of GDP devoted to healthcare from 17.3\% to 19.6\% and signaling rising financial pressure \citep{keehan2023national}. These trends underscore the importance of evaluating real-world costs of care using observational data. Such studies commonly involve continuous cost outcomes that are right-skewed and heavy-tailed, complicating prospective design when confounding adjustment is required. An illustrative example arises in analyses of post-progression costs in advanced non-small cell lung cancer (NSCLC) \citep{skinner2018healthcare}. NSCLC accounts for approximately 80\% of lung cancers, with half of patients diagnosed at stage IV \citep{travis2013new,siegel2018cancer}, and its high treatment costs highlight the tension between access to innovative therapies and healthcare sustainability \citep{jakovljevic2017disability}. We conduct a simulation case study designed to reflect the key statistical features of observational cost analyses in advanced NSCLC, including continuous, right-skewed outcomes, multiple baseline covariates, and confounding adjustment in real-world data.

We extend the simulation framework of Section~\ref{subsec:simulation-settings}
 to a continuous outcome with multiple baseline covariates and heteroskedastic heavy-tailed errors. Specifically, we generate i.i.d. observations $O=(X,B_1,B_2,B_3,T,Y)$, where $X\sim N(0,1)$ and $B_1$, $B_2$, $B_3$ are independently Bernoulli variables with probabilities 0.23, 0.11, and 0.54, respectively. Treatment assignment follows a logistic propensity score model:
\[
\Pr(T=1\mid X,B_1,B_2,B_3)
=
\expit\!\left\{\eta_0 + 0.6\left(0.7X+0.8B_1+1.0B_2+0.5B_3\right)\right\},
\]
where $\eta_0$ is calibrated within each dataset to achieve $\Pr(T=1)\approx 0.36$.

The outcome is generated from an additive mean model with covariate-dependent scale: 
\begin{align*}
\mu(X,B_1,B_2,B_3,T)
&=
18{,}000 + 1200X + 3500B_1 + 4500B_2 + 2000B_3 + \Delta T,\\
\sigma(X,B_1,B_2)&=3000\exp\!\left(0.15|X|+0.10B_1+0.15B_2\right),
\end{align*}
where $\Delta=1500$.
To reflect heavy-tailed healthcare cost data, errors follow a standardized Student-$t$ distribution with $\nu=4$:
\[
Y=\mu(X,B_1,B_2,B_3,T)+\sigma(X,B_1,B_2)\tilde\varepsilon,
\qquad \tilde\varepsilon\sim t_{\nu},\ \mathrm{Var}(\tilde\varepsilon)=1.
\]

The analysis model is the identity-link MSM $\mathbb{E}\{Y(t)\}=\beta_0+\beta_1 t$, so that $\beta_1$ represents the marginal mean difference in cost. All estimation and inference proceed via the sample IPTW stacked estimating equations described in Section~\ref{sec:methods}. For the RCT benchmark, $\rho=0.36$ and $n_{\text{pilot}}=350$ from \citet{skinner2018healthcare}---see Table~\ref{tab:rct_variance}.

\subsection{Results}

\begin{table}[t]
\centering
\caption{Continuous case study results across $R=1000$ independently generated pilots ($n_{\mathrm{pilot}}=350$ per pilot; $B=1000$ pilot-level bootstraps). For each stability choice, we report the induced prospective mean and median sample size across pilots. The hit rate is defined as the proportion of pilots for which the interpolated empirical power at the induced $n$ meets or exceeds $0.80$. Empirical power curves were estimated using 2000 Monte Carlo replicates per grid point. Reported power intervals correspond to empirical 95\% confidence intervals, defined as the 2.5th and 97.5th percentiles of the empirical power distribution across pilots. Proposed sample sizes are rounded up to the nearest integer.}
\label{tab:continuous_results}
\begin{tabular}{llccc}
\toprule 
Scenario & Stability choice & \shortstack{Prospective Sample Size\\ Mean (Median)} & \shortstack{Mean Power (\%) \\ (95\% CI)} & \shortstack{Hit Rate (\%) \\ (Power $\geq 80\%$)} \\
\midrule
\multicolumn{5}{l}{\textbf{Continuous cost outcome}}\\
& RCT benchmark & 314 (307) & 90.3 (85.6, 95.0) & 100.0 \\
& Median ($Q_{0.5}$) & 210 (197) & 78.1 (65.9, 94.3) & 38.7 \\
& UCB--Median & 213 (199) & 78.5 (66.2, 94.9) & 41.4 \\
& Mean & 218 (202) & 79.0 (66.2, 94.9) & 44.7 \\
& UCB--Mean & 221 (204) & 79.3 (66.9, 95.0) & 46.0 \\
& Quantile ($Q_{0.7}$) & 236 (217) & 81.2 (69.3, 95.1) & 58.0 \\
& Quantile ($Q_{0.9}$) & 284 (252) & 85.2 (72.9, 95.6) & 78.3 \\
\bottomrule
\end{tabular}
\end{table}

Table~\ref{tab:continuous_results} summarizes the prospective sample size distributions induced by the pilot and the empirical power calibration for the continuous outcome setting. Unlike the binary and count case studies, the RCT-style benchmark here appears conservative relative to the intended IPTW-MSM analysis. Although the RCT-based design ($n_{\mathrm{RCT,cont}}=314$) achieves mean empirical power of $90.3\%$ (95\% CI: 85.6\%--95.0\%), substantially exceeding the nominal $80\%$ target, this reflects variance miscalibration. Unlike the binary and count settings---where RCT formulas underestimate variability due to weighting---in this continuous setting, the benchmark variance formula overstates the variability of the IPTW estimator.

The pilot-based framework instead produces sample sizes that more closely align with the intended operating characteristics of the IPTW estimator. Central stability summaries such as the median and mean yield proposed designs in the range of approximately $200$--$220$, with corresponding mean power estimates near the $80\%$ target (78.1\%--79.3\%). As in the case studies, increasing the conservatism of the stability functional (e.g., moving to higher quantiles of the bootstrap LSVF distribution) improves reliability across pilot realizations. For example, moving from $Q_{0.5}$ to $Q_{0.7}$ raises the hit rate from $38.7\%$ to $58.0\%$ while maintaining mean power near the nominal target. Using the upper quantile $Q_{0.9}$ further improves calibration (hit rate $78.3\%$; mean power $85.2\%$) at the cost of moderate sample size inflation. To examine a setting where RCT-based formulas are appropriate, we conducted additional simulation with constant propensity scores; see Supplementary Appendix~\ref{app:S3}.

\section{Discussion}
\label{sec:discussion}

This paper proposes a pilot-based framework for prospective sample size determination when the primary analysis is an IPTW MSM estimated via stacked estimating equations. The central design quantity is the LSVF,
$V = n\,\mathrm{Var}(\widehat\beta_1)$, of the target causal estimator, evaluated under the same propensity score model, weighting scheme, and estimating equations intended for the definitive study. In observational settings, $V$ not only reflects the variability of the outcome, but also the realized covariate distribution, the prevalence of treatment, and the stability of the estimated inverse-probability weights. Consequently, RCT-style formulas that treat variance as primarily outcome-driven may misrepresent the variability of weighted estimators when applied to causal analyses.

A defining feature of the proposed approach is its reliance on pilot information to estimate the LSVF under a planned analysis. This reflects a general principle of prospective design: Power calculations inherently depend on prior knowledge or defensible assumptions about key features of the study, such as baseline risk, treatment prevalence, covariate distributions, and confounders. Here, pilot data---either observed individual-level data or synthetic data generated from a credible data-generating mechanism---are used to approximate the large-sample variance of the stacked estimator. If the pilot reasonably represents the target setting and standard causal assumptions hold, the resulting design explicitly targets the operating characteristics of the estimand-defined estimator, incorporating pilot evidence into confirmatory study planning.

Several consistent patterns emerge across the case studies. Mean-based stabilization is generally more conservative than median-based stabilization, reflecting its sensitivity to the occasional large LSVF realizations, whereas UCB-adjusted summaries introduce a modest additional upward shift in sample size that accounts for pilot uncertainty. The comparison with RCT-style benchmarks also varies across settings. In the binary and very rare count scenarios, RCT formulas underestimated the required sample size because weighting and limited covariate overlap inflate the variance of the causal estimator relative to the unweighted benchmark. In contrast, for the continuous outcome, the RCT-style benchmark was conservative and produced overpowered designs. These results indicate that discrepancies between RCT-style formulas and causal analyses depend on the relationship between outcome variability and weighting-induced variance inflation. Our findings suggest that stabilization should be viewed as a transparent design choice, informed by pilot diagnostics, study objectives, feasibility considerations, and risk tolerance for underpowered studies.

Our proposed framework enjoys broad applicability. Because the method is formulated entirely through stacked estimating equations and sandwich variance estimation, it applies to a wide array of outcome types within the stacked estimating-equation framework. Binary, count, and continuous endpoints are accommodated through appropriate link and variance choices, for which the pilot-based variance stabilization strategy remains unchanged. This estimator-aligned formulation contrasts with outcome-specific analytic power formulas that do not apply to different outcome types. Note that our framework is presented using an MSM targeting the population-average treatment effect, although alternative estimators (e.g., G-computation \citep{robins1986new} or doubly robust methods \citep{chernozhukov2018double}) could be incorporated within the same GEE framework, with additional complexity. Extending the framework to time-to-event outcomes via martingale-based estimating equations \citep{andersen1993statistical} is a natural direction for future work, but requires careful handling of censoring and specification of causal estimands.

Several limitations and potential extensions merit consideration. 
First, although the framework follows general principles of prospective design, its implementation requires either informative pilot data or a credible data-generating mechanism for synthetic pilot construction. These requirements inevitably rely on assumptions that may be unverifiable \citep{liu2025sample} or even unavailable (e.g., early-phase development or rare-disease space). Second, even when pilot data are available, the validity of the resulting design depends on the extent to which the pilot reflects the target study population and intended analysis strategy. Meaningful differences in covariate distributions, treatment prevalence, or weight stability between pilot and definitive study may lead to variance miscalibration. Finally, we focused on unstabilized IPTW and a working-independence MSM estimating equation. Alternative weighting strategies (e.g., stabilized weights, truncation \citep{crump2009dealing,sturmer2021propensity}, or overlap weights \citep{li2019addressing}) and alternative working variance specifications may alter the behavior of variance estimators. These extensions can be accommodated directly within the proposed framework by redefining the stacked estimating equations used consistently at both the pilot and design stages; however, their detailed investigation is beyond the scope of this paper and deferred to future work.

\section*{Disclaimer}
The views expressed in this article should not be construed to represent those of the U.S. Food and Drug Administration.

\section*{Conflict of Interests}
The authors declare no conflicts of interest.

\section*{Data Availability Statement}
The simulation code is available on GitHub at \url{https://github.com/tkh5956/Samplesize_IPTW}.

    {
    \setlength{\bibsep}{0pt}
    \bibliographystyle{unsrtnat}
    \bibliography{bibtex}
    }

\clearpage
\setcounter{section}{0}
\renewcommand*{\thetable}{\Alph{table}}
\renewcommand*{\thefigure}{\Alph{figure}}
\renewcommand*{\thesection}{\Alph{section}}

\newpage
\section{Supplementary Appendix: Pseudocode}
\label{app:S1}

\begin{algorithm}[ht]
\caption{Bootstrap-Stabilized Design Variance Selection from Pilot Data}
\label{alg:stable_design_variance}

\KwIn{Pilot dataset $\mathcal{D}_{\mathrm{pilot}}$ of size $n_{\mathrm{pilot}}$; number of pilot bootstraps $B$; stability functional $\mathcal{F}$; generic functional $\phi$; number of second-stage bootstraps $B_{\mathrm{ucb}}$; upper-tail level $\gamma_{\mathrm{ucb}}$}
\KwOut{Stable design variance estimate $\widehat V_{\mathrm{stable}}$ and upper confidence bound $\widehat V_{\mathrm{stable}}^{\mathrm{UCB}}$}

Apply the stacked estimating procedure to $\mathcal{D}_{\mathrm{pilot}}$\;
Extract $\widehat{\mathrm{Var}}_{\mathrm{pilot}}(\widehat\beta_1)$ from the $\beta$-block of~\eqref{eq:empirical_sandwich}\;
Compute
\[
\widehat V_{\mathrm{pilot}}
=
n_{\mathrm{pilot}}\widehat{\mathrm{Var}}_{\mathrm{pilot}}(\widehat\beta_1)
\]

\For{$b=1,\ldots,B$}{
    Draw bootstrap resample $\mathcal{D}_{\mathrm{pilot}}^{*(b)}$ from $\mathcal{D}_{\mathrm{pilot}}$\;
    Apply the stacked estimating procedure to $\mathcal{D}_{\mathrm{pilot}}^{*(b)}$
    Compute
    \[
    \widehat V^{*(b)}
    =
    n_{\mathrm{pilot}}\widehat{\mathrm{Var}}^{*(b)}(\widehat\beta_1)
    \]
}

Use $\{\widehat V^{*(b)}\}_{b=1}^B$ as a proxy for the sampling distribution of $V$\;

Compute the stable design value
\[
\widehat V_{\mathrm{stable}}
=
\mathcal{F}\!\left(
\widehat V^{*(1)},\ldots,\widehat V^{*(B)}
\right)
\]

\Indp
For example, $\mathcal{F}=Q_q(\cdot)$ for $q\in\{0.5,0.7,0.9\}$, or $\mathcal{F}=\mathrm{mean}(\cdot)$
\Indm

\For{$k=1,\ldots,B_{\mathrm{ucb}}$}{
    Resample with replacement indices $\pi_k(1),\ldots,\pi_k(B)$ from $\{1,\ldots,B\}$\;
    Compute
    \[
    \phi^{*(k)}
    =
    \phi\!\left(
    \widehat V^{*\pi_k(1)},\ldots,\widehat V^{*\pi_k(B)}
    \right)
    \]
}

Compute the bootstrap-based upper confidence bound
\[
\widehat V_{\mathrm{stable}}^{\mathrm{UCB}}
=
Q_{1-\gamma_{\mathrm{ucb}}}\!\left(
\phi^{*(1)},\ldots,\phi^{*(B_{\mathrm{ucb}})}
\right)
\]

\Return{$\widehat V_{\mathrm{stable}},\ \widehat V_{\mathrm{stable}}^{\mathrm{UCB}}$}\;

\end{algorithm}
\newpage

\section{Supplementary Appendix: Conceptual illustration of the two-level bootstrap procedure}
\label{app:S2}

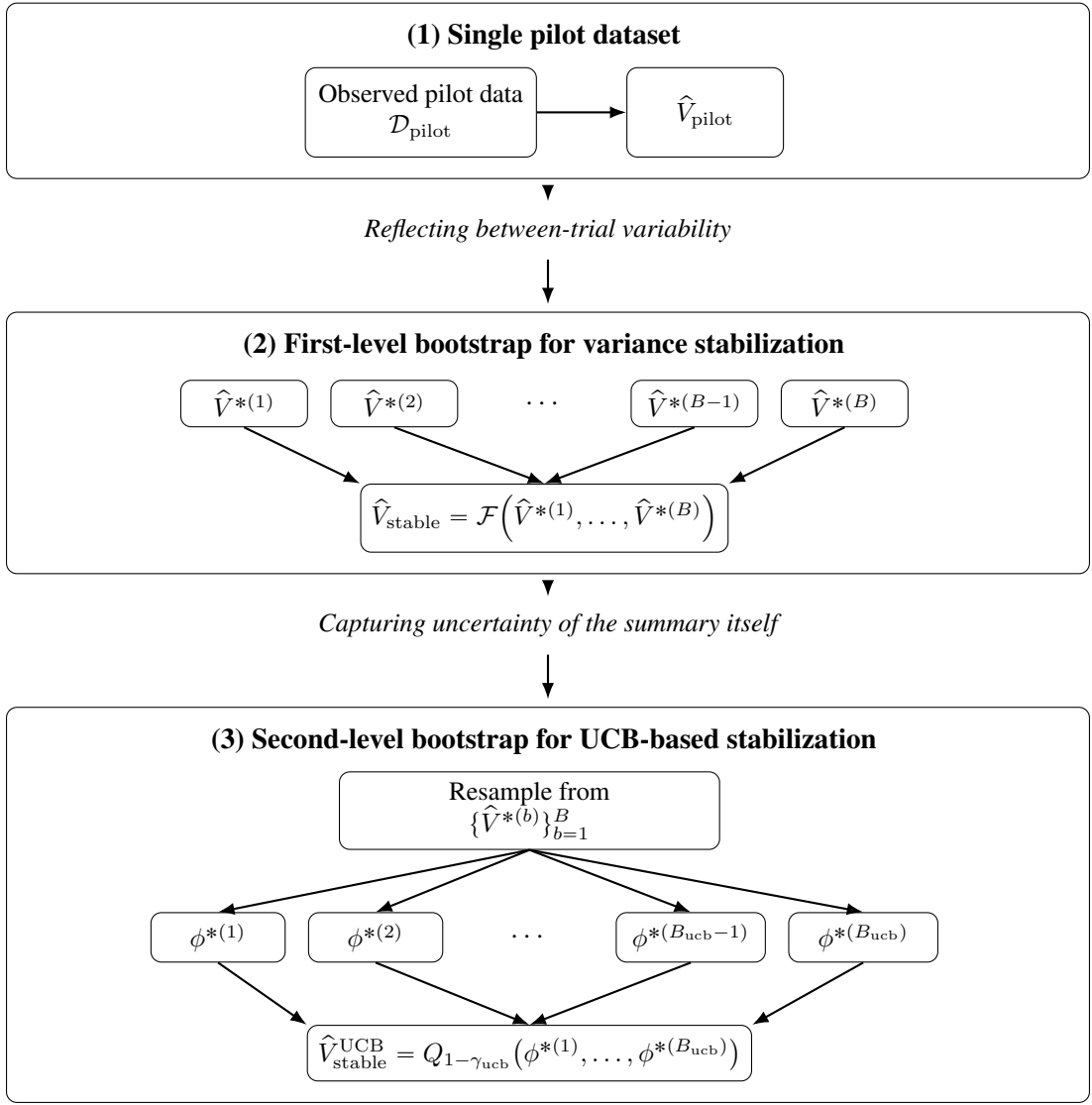
\begin{figure}[h]
\centering
\begin{tikzpicture}[
  node distance=10mm and 8mm,
  panel/.style={draw, rounded corners, align=center, inner sep=8pt, minimum width=0.88\linewidth},
  smallbox/.style={draw, rounded corners, align=center, inner sep=4pt, font=\small},
  arrow/.style={-{Latex[length=2.2mm]}, thick},
  lab/.style={font=\small\itshape, align=center}
]

\node[panel] (p1) {
\begin{minipage}{0.82\linewidth}
\centering
\textbf{(1) Single pilot dataset}\\[2mm]

\begin{tikzpicture}[baseline=(current bounding box.center)]
\node[smallbox, minimum width=3.1cm, minimum height=1.2cm] (pilot) {Observed pilot data\\$\mathcal{D}_{\mathrm{pilot}}$};
\node[smallbox, right=1.2cm of pilot, minimum width=2.1cm, minimum height=1.2cm] (vhat) {$\widehat V_{\mathrm{pilot}}$};
\draw[arrow] (pilot) -- (vhat);
\end{tikzpicture}
\end{minipage}
};

\node[lab, below=4mm of p1] (l12) {Reflecting between-trial variability};
\draw[arrow] ([yshift=-1mm]p1.south) -- ([yshift=1mm]l12.north);

\node[panel, below=8mm of l12] (p2) {
\begin{minipage}{0.82\linewidth}
\centering
\textbf{(2) First-level bootstrap for variance stabilization}\\[2mm]

\begin{tikzpicture}[baseline=(current bounding box.center)]
\node[smallbox, minimum width=1.7cm] (v1) {$\widehat V^{*(1)}$};
\node[smallbox, right=3mm of v1, minimum width=1.7cm] (v2) {$\widehat V^{*(2)}$};
\node[right=3mm of v2, minimum width=1.7cm] (dots) {$\cdots$};
\node[smallbox, right=3mm of dots, minimum width=1.7cm] (vb1) {$\widehat V^{*(B-1)}$};
\node[smallbox, right=3mm of vb1, minimum width=1.7cm] (vb) {$\widehat V^{*(B)}$};

\node[smallbox, below=7mm of dots, minimum width=3.2cm] (f) {$\widehat V_{\mathrm{stable}}=\mathcal{F}\!\left(\widehat V^{*(1)},\ldots,\widehat V^{*(B)}\right)$};

\draw[arrow] (v1.south) -- (f.north west);
\draw[arrow] (v2.south) -- (f.north);
\draw[arrow] (vb1.south) -- (f.north);
\draw[arrow] (vb.south) -- (f.north east);
\end{tikzpicture}
\end{minipage}
};

\draw[arrow] ([yshift=-1mm]l12.south) -- ([yshift=1mm]p2.north);

\node[lab, below=4mm of p2] (l23) {Capturing uncertainty of the summary itself};
\draw[arrow] ([yshift=-1mm]p2.south) -- ([yshift=1mm]l23.north);

\node[panel, below=8mm of l23] (p3) {
\begin{minipage}{0.82\linewidth}
\centering
\textbf{(3) Second-level bootstrap for UCB-based stabilization}\\[2mm]

\begin{tikzpicture}[baseline=(current bounding box.center)]
\node[smallbox, minimum width=1.8cm] (phi1) {$\phi^{*(1)}$};
\node[smallbox, right=3mm of phi1, minimum width=1.8cm] (phi2) {$\phi^{*(2)}$};
\node[right=3mm of phi2, minimum width=1.7cm] (dots2) {$\cdots$};
\node[smallbox, right=3mm of dots2, minimum width=2.0cm] (phik1) {$\phi^{*(B_{\mathrm{ucb}}-1)}$};
\node[smallbox, right=3mm of phik1, minimum width=2.0cm] (phik) {$\phi^{*(B_{\mathrm{ucb}})}$};

\node[smallbox, above=8mm of dots2, minimum width=5.1cm] (resample) {Resample from\\$\{\widehat V^{*(b)}\}_{b=1}^B$};

\node[smallbox, below=8mm of dots2, minimum width=5.1cm] (ucb) {$\widehat V_{\mathrm{stable}}^{\mathrm{UCB}}
=
Q_{1-\gamma_{\mathrm{ucb}}}\!\left(\phi^{*(1)},\ldots,\phi^{*(B_{\mathrm{ucb}})}\right)$};

\draw[arrow] (resample.south) -- (phi1.north);
\draw[arrow] (resample.south) -- (phi2.north);
\draw[arrow] (resample.south) -- (phik1.north);
\draw[arrow] (resample.south) -- (phik.north);
\draw[arrow] (phi1.south) -- (ucb.north west);
\draw[arrow] (phi2.south) -- (ucb.north);
\draw[arrow] (phik1.south) -- (ucb.north);
\draw[arrow] (phik.south) -- (ucb.north east);
\end{tikzpicture}
\end{minipage}
};

\draw[arrow] ([yshift=-1mm]l23.south) -- ([yshift=1mm]p3.north);

\end{tikzpicture}
\caption{Conceptual illustration of the two-level bootstrap procedure for UCB-based variance stabilization. Starting from a single pilot dataset, the first-level bootstrap reflects between-trial variability in the unit variance estimate, while the second-level bootstrap captures uncertainty in the summary functional itself and yields a data-adaptive upper confidence bound.}
\label{fig:ucb_bootstrap_concept}
\end{figure}

\newpage

\section{Supplementary Appendix: Simulation under constant propensity (no confounding)}
\label{app:S3}
To examine a setting in which RCT-based formulas are appropriate, we conducted a supplementary analysis with constant propensity, where treatment assignment was independent of baseline covariates and the propensity model was intercept-only. In this setting, the RCT benchmark achieved mean power close to the nominal $80\%$ target across scenarios in contrast to the case studies in the main manuscript. The proposed pilot-based approach likewise showed robust performance: central summaries yielded mean power near $80\%$, while higher-quantile stabilization improved hit rates at the cost of modest sample size inflation. These results confirm that the proposed framework remains valid when conventional RCT formulas are appropriate, while retaining its flexibility to accommodate more complex observational settings.

\begin{table}[h]
\centering
\caption{(Constant propensity) Binary case study results across $R=1000$ independently generated pilots ($n_{\mathrm{pilot}}=600$ per pilot; $B=1000$ pilot-level bootstraps). For each stability choice, we report the induced prospective mean and median sample size across pilots. The hit rate is defined as the proportion of pilots for which the interpolated empirical power at the induced $n$ meets or exceeds $0.80$. Empirical power curves were estimated using 2000 Monte Carlo replicates per grid point. Reported power intervals correspond to empirical 95\% confidence intervals, defined as the 2.5th and 97.5th percentiles of the empirical power distribution across pilots. Proposed sample sizes are rounded up to the nearest integer.}
\label{tab:binary_results_rct}
\begin{tabular}{llccc}
\toprule 
Scenario & Stability choice & \shortstack{Prospective Sample Size\\ Mean (Median)} & \shortstack{Mean Power (\%) \\ (95\% CI)} & \shortstack{Hit rate (\%) \\(Power$\geq 80\%$)} \\
\midrule
\multicolumn{5}{l}{\textbf{(A) Low event rate: $p_0=0.03$}}\\
& RCT benchmark     & 1940 (1940) & 78.4 (78.4, 78.4) & 0.0\\
& Median ($Q_{0.5}$)   & 2294 (2162) & 81.7 (65.0, 95.9) & 57.7 \\
& UCB--Median       & 2335 (2203) & 82.3 (65.7, 96.2) & 61.0 \\
& Mean              & 2527 (2346) & 84.3 (67.4, 98.2) & 69.8 \\
& UCB--Mean         & 2578 (2346) & 84.9 (68.1, 98.2) & 71.2 \\
& Quantile ($Q_{0.7}$) & 2686 (2481) & 86.2 (70.4, 98.2) & 78.5 \\
& Quantile ($Q_{0.9}$) & 3579 (3146) & 91.8 (78.6, 98.7) & 95.6 \\
\midrule
\multicolumn{5}{l}{\textbf{(B) Higher event rate: $p_0=0.10$}}\\
& RCT benchmark     & 682 (682) & 79.4 (79.4, 79.4) & 0.0 \\
& Median ($Q_{0.5}$)   & 712 (702) & 79.4 (72.0, 87.3) & 43.4 \\
& UCB--Median       & 717 (707) & 79.7 (72.1, 87.4) & 45.9 \\
& Mean              & 724 (713) & 80.0 (72.2, 87.5) & 49.3 \\
& UCB--Mean         & 729 (717) & 80.2 (72.4, 87.6) & 51.9 \\
& Quantile ($Q_{0.7}$) & 757 (745) & 81.6 (73.3, 88.4) & 64.8 \\
& Quantile ($Q_{0.9}$) & 838 (821) & 84.6 (75.9, 91.1) & 87.9 \\
\bottomrule
\end{tabular}
\end{table}

\begin{table}[h]
\centering
\caption{(Constant propensity) Count case study results across $R=1000$ independently generated pilots ($n_{\mathrm{pilot}}=5000$ per pilot; $B=1000$ pilot-level bootstraps). For each stability choice, we report the induced prospective sample size summarized as mean (median) across pilots. The hit rate is defined as the proportion of pilots for which the interpolated empirical power at the induced $n$ meets or exceeds $0.80$. Empirical power curves were estimated using 2000 Monte Carlo replicates per grid point. Reported power intervals correspond to empirical 95\% confidence intervals (2.5th and 97.5th percentiles across pilots). Proposed sample sizes are rounded up to the nearest integer.}
\label{tab:count_results_rct}
\begin{tabular}{llccc}
\toprule 
Scenario & Stability choice & \shortstack{Prospective Sample Size\\ Mean (Median)} & \shortstack{Mean Power (\%) \\ (95\% CI)} & \shortstack{Hit Rate (\%) \\ (Power $\geq 80\%$)} \\
\midrule
\multicolumn{5}{l}{\textbf{Count outcome: $\lambda_0=0.008$}}\\
& RCT benchmark & 12284 (12284) & 81.7 (81.7, 81.7) & 100.0 \\
& Median ($Q_{0.5}$) & 13285 (12633) & 82.2 (69.2, 95.7) & 58.7 \\
& UCB--Median & 13481 (12849) & 82.7 (69.5, 96.1) & 61.2 \\
& Mean & 14125 (13333) & 83.9 (70.1, 97.6) & 67.7 \\
& UCB--Mean & 14344 (13520) & 84.4 (70.6, 97.6) & 69.5 \\
& Quantile ($Q_{0.7}$) & 15094 (14161) & 86.0 (73.6, 97.6) & 79.8 \\
& Quantile ($Q_{0.9}$) & 18818 (17428) & 91.2 (77.6, 97.9) & 95.5 \\
\bottomrule
\end{tabular}
\end{table}

\begin{table}[h]
\centering
\caption{(Constant propensity) Continuous case study results across $R=1000$ independently generated pilots ($n_{\mathrm{pilot}}=350$ per pilot; $B=1000$ pilot-level bootstraps). For each stability choice, we report the induced prospective mean and median sample size across pilots. The hit rate is defined as the proportion of pilots for which the interpolated empirical power at the induced $n$ meets or exceeds $0.80$. Empirical power curves were estimated using 2000 Monte Carlo replicates per grid point. Reported power intervals correspond to empirical 95\% confidence intervals, defined as the 2.5th and 97.5th percentiles of the empirical power distribution across pilots. Proposed sample sizes are rounded up to the nearest integer.}
\label{tab:continuous_results_rct}
\begin{tabular}{llccc}
\toprule 
Scenario & Stability choice & \shortstack{Prospective Sample Size\\ Mean (Median)} & \shortstack{Mean Power (\%) \\ (95\% CI)} & \shortstack{Hit Rate (\%) \\ (Power $\geq 80\%$)} \\
\midrule
\multicolumn{5}{l}{\textbf{Continuous cost outcome}}\\
& RCT benchmark & 301 (295) & 81.4 (73.6, 90.7) & 62.4 \\
& Median ($Q_{0.5}$) & 287 (280) & 79.4 (69.7, 90.3) & 47.0 \\
& UCB--Median & 290 (282) & 79.8 (69.9, 90.9) & 49.4 \\
& Mean & 290 (283) & 79.8 (69.9, 91.2) & 49.5 \\
& UCB--Mean & 293 (284) & 80.0 (70.1, 91.2) & 50.6 \\
& Quantile ($Q_{0.7}$) & 309 (299) & 81.8 (72.0, 91.6) & 65.0 \\
& Quantile ($Q_{0.9}$) & 344 (329) & 84.8 (74.3, 92.8) & 84.7 \\
\bottomrule
\end{tabular}
\end{table}
\end{document}